\documentclass{article}

\begin{document}

\title{$\Lambda$-symmetries of Dynamical Systems, \\
Hamiltonian and Lagrangian equations}

\author{Giampaolo Cicogna\\ {\it Dipartimento di Fisica, Universit\`a di
Pisa}  \\
and {\it INFN, Sezione di Pisa,} \\
{\it Largo B. Pontecorvo 3, 50127 Pisa (Italy)} \\ {\tt
cicogna@df.unipi.it}}

\date{}

\maketitle

\newtheorem{theorem}{Theorem}
\newtheorem{example}{Example}
\newtheorem{corollary}{Corollary}

\begin{abstract} After a brief survey  of the definition and the properties of  
$\Lambda$-symmetries in the general context of dynamical systems, 
the notion of  ``$\Lambda$-constant of motion'' for Hamiltonian 
equations is introduced. If the Hamiltonian problem is derived from  
a $\Lambda$-invariant Lagrangian, it is shown how the Lagrangian 
$\Lambda$-invariance can be transferred into the Hamiltonian 
context and shown that the Hamiltonian equations turn out to be 
$\Lambda$-symmetric. Finally, the ``partial'' (Lagrangian) reduction 
of the Euler-Lagrange equations is compared with the reduction obtained 
for the corresponding Hamiltonian equations.
\end{abstract}

\section{Introduction ($\lambda$-symmetries)}

Let me briefly recall for the reader's convenience the basic definition of 
$\lambda$-symmetry (with lower case $\lambda$), originally introduced by C. 
Muriel and J.L. Romero in 2001 \cite{Cic:MR1,Cic:MR2}. 

Let me consider the simplest case of  a single ODE
$\Delta(t,u(t),\dot u,\ddot u\ldots)=0$ for the unknown function
$u\,=\,u(t)$ 
(I will denote by $t$ the independent variable, with the only exception 
of the final Section 4, because the applications 
I am going to propose will concern the case of Dynamical Systems (DS),
where  the independent variable is precisely the time $t$, and $\dot 
u={\rm d}u/{\rm d} t$, etc.). Given a vector field 
\[ X\,=\,\varphi(u,t){\frac {\partial} {\partial u}}+
\tau(u,t){\frac \partial {\partial t}}\]
the idea  is to suitably {\it modify} its prolongation rules. The first 
$\lambda$-prolongation $X^{(1)}_\lambda$ is the defined by
\begin{equation}\label{Cic:la1}X^{(1)}_\lambda\,=\, X^{(1)}+
\lambda (\varphi-\tau\dot u){\frac \partial {\partial \dot u}}\end{equation}
where  $\lambda=\lambda(u,\dot u,t)$ is a $C^\infty$ function, and 
$X^{(1)}$ is the standard first prolongation.
Other modifications have to be introduced for higher prolongations,
but in the present paper I need only just the first one.

\smallskip
An $n$-th order ODE $\Delta=0$ is said to be $\lambda$-invariant under 
$X^{(n)}_\lambda$ if
\[X^{(n)}_\lambda\Delta\Big|_{\Delta=0}\,=\,0\]
where $X^{(n)}_\lambda$ is the $n$-th $\lambda$-prolongation of $X$.

\smallskip
It should be emphasized that $\lambda$-symmetries are not properly 
symmetries, because they do not transform in general solutions of a 
$\lambda$-invariant equation into solutions, nevertheless
they share with standard Lie point-symmetries some important properties, namely: 
if an equation is $\lambda$-invariant, then

\smallskip
\noindent $\bullet$ the order of the equation can be lowered by one 

\smallskip
\noindent $\bullet$ invariant solutions can be found  
(notice that conditional symmetries do the same, but $\lambda$-symmetries  
are clearly {\it not} conditional symmetries)

\smallskip
\noindent $\bullet$ convenient new (``symmetry adapted'') variables can be 
suggested.

\smallskip
In the context of DS, which is the main 
object of this paper, the first two properties are not effective, the
third one is instead one of my starting points.

\smallskip
Before considering the role of $\lambda$-symmetries in DS, let me recall that 
many applications and extensions of this notion have been proposed in 
these 10 years: these include extensions to systems of ODE's, to PDE's, 
applications to variational principles and Noether-type theorems, the
analysis of their connections with nonlocal symmetries, with 
symmetries of exponential type, with hidden, or ``lost'' symmetries,
with potential, telescopic symmetries as well. Other investigations concern 
their  deep geometrical interpretation, with the introduction of a suitable notion of 
deformed Lie 
differential operators, the study of their dynamical effects in terms of
changes of reference frames, and so on.
Only the papers more directly involved with the argument considered in 
this paper will be quoted; for a fairly complete list of references
see e.g. \cite{Cic:CHam,Cic:Gatw,Cic:GC09}.
A very recent application concerns discrete difference equations 
\cite{Cic:LRdde}.

\section{$\Lambda$-symmetries for DS}

I am going to consider the case of dynamical systems, i.e. systems of 
first-order ODE's
\[\dot u_a\,=\,f_a(u,t)\quad\quad\ (a=1,\ldots, m)\]
for the $m>1$ unknowns $u_a\,=\,u_a(t)$.
 
Let me start with a  trivial (but significant) case: if the DS admits a rotation 
symmetry, then it is 
completely natural to introduce as new variables the radius $r$ and the 
angle $\theta$, and the DS immediately takes a simplified form, as well known.
However, in general, symmetries of DS may be very singular, and/or difficult 
to detect. An example can be useful: the DS
\[\dot u_1\,=\,u_1u_2 \quad\quad\ \dot u_2\,=\,-u_1^2\]
admits the (not very useful or illuminating) symmetry generated by 
(with $r^2=u_1^2+u_2^2$)
\[X\,=\,\Big({\frac {2u_1} {r^2}}-
{\frac {u_1u_2} {r^3}}\log{\frac {u_2-r} {{u_2+r}}}\Big)
{\frac {\partial} {\partial u_1}}+ 
\Big({\frac {2u_2}{r^2}-{\frac{u_1^2}{r^3}}
\log{\frac {u_2-r} {{u_2+r}}}\Big) {\frac {\partial} {\partial 
u_2}}}\ .\]
In this example the rotation (with a commonly accepted abuse of language, 
the same symbol $X$ denotes both the symmetry and its Lie generator)
\[X\,=\,u_2{\frac {\partial} {\partial u_1}}-u_1{\frac \partial {\partial u_2}}\]
is a {\it $\lambda$-symmetry}\ (its precise definition will be given in a 
moment), and {\it not} a symmetry in the ``standard''  sense; 
{\it nevertheless}, still introducing the variables as before, i.e.
$r$ and  $\theta$, the DS takes the very simple form
\[\dot r\,=\,0\quad\quad\ \dot\theta\,=\,-r\cos\,\theta\ .\]
This is just a first, simple example of the possible role of 
$\lambda$-symmetries in the context of DS.

\bigskip
\subsection{$\Lambda$-symmetries of general DS}

The natural way to extend the definition (\ref{Cic:la1}) of the first 
$\lambda$-prolongation of the vector field
\[X\,=\,\varphi_a(u,t){\frac \partial {\partial u_a}}+
\tau(u,t){\frac\partial {\partial t}}\,=\,
\varphi\cdot{\nabla_u}+\tau\partial_t\]
to the case of $m>1$ variables $u_a$ is the following (sum over repeated 
indices)
\[X_\Lambda^{(1)}\,=\, X^{(1)}+\Lambda_{ab}(\varphi_b-\tau\dot u_b)
\cdot\nabla_{\dot u_a}\] 
where now $\Lambda=\Lambda(t,u_a,\dot u_a)$  is a $(m\times m)$ matrix; 
accordingly, I
denote by the upper case $\Lambda$ these symmetries in this context.

To simplify, let me assume from now on $\tau\,=\,0$ 
(or use evolutionary vector field, it is not restrictive). 

Then the given DS is $\Lambda$-invariant under $X$ (or $X$ is a 
$\Lambda$-symmetry for the DS), i.e. $X_\Lambda^{(1)}(\dot u-f)|_{\dot u=f}=0$, 
if and only if
\[
[\,f,\varphi\,]\,_a+\partial_t\varphi_a\,=\,-(\Lambda\varphi)_a \quad\qquad 
(a=1,\ldots,m)\, \]
where 
\[ [\,f, \varphi\,]\,_a\equiv f_b\nabla_{u_b}\varphi_a-\varphi_b
\nabla_{u_b}f_a \ .\]

Given $X$, we now introduce  the following new $m+1$ ``canonical'' (or  
{\it symmetry-adapted}) 
variables ({\it notice that they are independent of} $\Lambda$): precisely, 
$m-1$ variables $w_j=w_j(u)$ which, together with the time $t$, are 
$X$- invariant:
\[X\,w_j\,=\,X\,t\,=\,0\quad\quad (j=1,\ldots,m-1)\]
and  the coordinate $z$, ``rectifying''  the action of $X$, i.e.
\[ X\,=\,{\frac \partial {\partial z}}\ .\]
Writing the given DS in these new variables, we obtain a ``reduced'' form of the 
DS, as stated by the following theorem \cite{Cic:Olv,Cic:MRVi,Cic:CLa}.

\begin{theorem}
Let $X$ be a $\Lambda$-symmetry for a given DS; once the DS is written in terms 
of the  new variables $w_j,z,t$, i.e. 
\[ \dot w_j  \,=\,W_j(w,z,t) \quad\quad\  \dot z \,=\,Z(w,z,t)\] 
the dependence on $z$ of the r.h.s. 
$W_j\,,Z$ is controlled by the formulas
\[{\frac{\partial W_j}{\partial z}}\,=\,
{\frac{\partial w_j}{\partial u_a}}(\Lambda\varphi)_a\equiv M_j 
\quad\quad\
{\frac{\partial Z}{\partial z}}\,=\,
{\frac{\partial z}{\partial u_a}}(\Lambda\varphi)_a\ \equiv M_{m}\ .\]
One has:
\[\bullet\  \ If \  \Lambda=0\  \ then\ M_j=M_m=0 \ \ and \ \
W_j\,,Z \ are\ independent\ of \ z
\hskip 7.6cm\]
\[\bullet\ \  If \  \Lambda\,=\,\lambda I\  \ then\   only\  Z\   depends \
on\  z   \hskip8.7cm\]

\vskip.05cm\noindent
$\bullet$ Otherwise,  a  ``partial''  reduction\ is\ obtained:\
If some  $M_k=0$, then  $W_k$  is independent of $z$. 
In terms of the new variables, the $\Lambda$-prolongation becomes
\[  X_{\Lambda}^{(1)}\,=\,{\frac\partial {\partial 
z}}+M_{j}{\frac\partial {\partial\dot w_j}}+M_{m}{\frac\partial 
{\partial\dot z}}\ .\]
\end{theorem}

The first case ($\Lambda=0$) clearly means that $X$ is an {\it exact}, or 
standard Lie point-symmetry \cite{Cic:Olv}; the second one has been considered 
in detail by Muriel 
and Romero \cite{Cic:MRVi} (notice that actually  it would be enough to 
require $\Lambda\varphi=\lambda\varphi$); the last case has been dealt 
with in \cite{Cic:CLa}:
several situations can be met, depending on the number of vanishing 
$M_j$  (e.g., one may obtain triangular DS, or similar).

\subsection{Hamiltonian DS}

I now consider the special case in which the DS is a {\it Hamiltonian} DS.
Obvious changes in the notations can be introduced: the $m$ variables 
$u=u_a(t)$ are replaced by the $m=2n$ variables $q_\alpha(t),p_\alpha(t)\ 
(\alpha=1,\ldots,n)$,
and the DS is now the system of the Hamiltonian equations of motions for the 
given Hamiltonian $H=H(q,p,t)$:
\[ \dot u\,=\,J\nabla H\equiv F(u,t)\quad\,,\quad \ \nabla\equiv 
\nabla_u\equiv(\nabla_q,\nabla_p)\]
where $J$ is the standard symplectic matrix
\[J\,=\,\pmatrix{0&I \cr  -I&0}\ . \]
A vector field\ $X$ can be written accordingly (with $a=1,\ldots,2n\,;\,
\alpha=1,\ldots,n$)
\[X=\varphi_\alpha(u,t){\frac\partial {\partial q_\alpha}}+
\psi_\alpha (u,t){\frac\partial {\partial 
p_\alpha}}\equiv{\bf \Phi}\cdot\nabla_u\quad\,,\quad\ {\bf
\Phi}\equiv(\varphi_\alpha,\psi_\alpha)\]
and all the above discussion clearly holds if $X$ is a $\Lambda$-symmetry 
for an Hamiltonian DS. Clearly, 
here  $\Lambda$ is a $(2n\times 2n)$ matrix. But Hamiltonian problems possess 
certainly a {\it richer} structure with respect to general DS, which deserves 
to be exploited; a first  instance is clearly provided by the notion of 
conservation rules, with its related topics.

Let me then distinguish two cases:

\smallskip\noindent
$(i$) $X$ admits a {\it generating function} $G(u,t)$ 
(then $X$ is often called a ``Hamiltonian symmetry''):
\begin{equation}\label{Cic:Phi}{\bf \Phi}\,=\,J\nabla G\quad{\rm i.e.}\quad  
\varphi\,=\,\nabla_pG \ ,\  \psi\,=\,-\nabla_qG
\end{equation}
this implies $\nabla D_tG=0$, where $D_t$ is the total derivative,
i.e. $G$ is a constant of motion, $D_tG=0$, possibly apart from an additional 
time-dependent term, as well known.

\smallskip\noindent
$(ii$) $X$ does not admit a generating function: also in this case, defining
\begin{equation}\label{Cic:defS}S(u,t)\equiv \nabla\cdot{\bf \Phi}\quad one 
\ has \ that\  \quad D_tS\,=\,0
\end{equation}
and therefore, if $S\not=$ const, then $S$ is a first integral  (the examples 
known to me of first integrals of this form are rather tricky, 
being usually obtained multiplying symmetries  by first integrals; but
they ``in principle'' exist, and their presence will be important for the
following discussion, see subsect. 3.4).

\smallskip

Direct calculations can show the following:

\begin{theorem}

If the Hamiltonian equations of motion admit a  
$\,\Lambda$-symmetry $X$ with a matrix $\Lambda$, then:

\noindent
in case $(i)$ 
\[\nabla(D_tG)\,=\,J\,\Lambda\,{\bf \Phi}\,=\,J\,\Lambda\,J\,\nabla\,G\]
\noindent
in case $(ii)$ 
\[D_tS\,=\,-\nabla(\Lambda\,{\bf \Phi})\ .\]
\end{theorem}

\smallskip

When this happens, $G$ (resp. $S$) will be called a 
``{\it $\Lambda$-constant of motion}''.

\bigskip
If $\Lambda\!=\!0$, i.e. when $X$ is a ``standard'' (or ``exact'') symmetry, 
the above equations become clearly the usual conservation rules; 
$\Lambda$-symmetries can then be viewed as ``perturbations'' of  the exact 
symmetries. More explicitly, the equations in Theorem 2 state the precise
``deviation'' from the conservation of $G$ (resp. of $S$)  
due to the fact that the invariance under $X$ is ``broken'' by the 
presence of a nonzero matrix $\Lambda$.

As a special case for case $(i)$, the following Corollary may be of interest:
\smallskip
\begin{corollary}
Under mild assumptions ($\Lambda\,{\bf \Phi}\!=\!\lambda\,{\bf \Phi}$, 
$\lambda\!=\!\lambda(G)$),
then the $\Lambda$-constant of motion $G$  satisfies a  ``completely 
{\it separated} equation'', involving only $G(t)$:
\[\dot G\,=\,\gamma(t,G)\ . \]
\end{corollary}
This equation expresses how much the conservation of $G(t)$ is ``violated'' 
along the time evolution. If $\Lambda$ is in some sense ``small'', then 
$G$ is ``almost'' conserved.

\section{When a $\Lambda$-symmetry of the Hamiltonian equations is 
inherited  by a $\Lambda$-invariant  Lagrangian}

\subsection{$\Lambda$-invariant  Lagrangians, Noether theorem and\\
$\Lambda$-conservation rules}

Let me consider (for simplicity) only first-order Lagrangians:
\[{{\cal L}}\,=\,{{\cal L}}(q_\alpha,\dot q_\alpha,t)\quad\quad\quad 
(\alpha=1,\ldots,n) \]
Such a Lagrangian is said to be 
$\Lambda^{({\cal L})}$-invariant\cite{Cic:MRO,Cic:CGNoe} 
under  \[ X^{({{\cal L}})}\,=\,
\varphi_\alpha(q,t){\frac\partial {\partial q_\alpha}}\,=\,
\varphi\cdot\nabla_q\]
if there is an $(n\times n)$ matrix 
\[\Lambda^{({{\cal L}})}\,=\,\Lambda^{({{\cal L}})}(q,\dot q,t)\] 
such that 
\[\Big(X_\Lambda^{({{\cal L}})}\Big)^{(1)}({{\cal L}})\,=\,0\]
where  $\Big(X_\Lambda^{({{\cal L}})}\Big)^{(1)}$ is the first
$\Lambda^{({{\cal L}})}$-prolongation  of $X^{({{\cal L}})}$ 
(the notation is rather heavy, to carefully distinguish the Lagrangian 
case from the Hamiltonian one, to be considered in the next subsection).

We then have \cite{Cic:CGNoe}

\begin{theorem}
If the Lagrangian ${{\cal L}}$ is $\Lambda^{({{\cal L}})}$-invariant
under $X^{({{\cal L}})}$  then, putting
\[{{\cal P}}_{\alpha\beta}\,=\,\varphi_\alpha p_\beta \quad\quad with 
\quad\quad p_\beta\,=\,
{\frac{\partial{{\cal L}}}{\partial \dot q_\beta}}\]
one has
\[ D_t{\bf P}=
-\Lambda^{({{\cal L}})}_{\alpha\beta}\varphi_\beta{\frac{\partial{{\cal L}}}{\partial 
\dot q_\alpha}} \,=\, -\big(\Lambda^{({{\cal L}})}\varphi\big)_\alpha p_\alpha\]
where ${\bf P}={{\tt Tr}}({{\cal P}})\,=\,\varphi_\alpha p_\alpha$;
or also, introducing a ``deformed derivative'' ${\widehat D}_{t}$
\[({\widehat D}_{t})_{\alpha\beta}\equiv D_t\delta_{\alpha\beta}+
\Lambda^{({{\cal L}})}_{\alpha\beta} \quad 
then \quad {{\tt Tr}}({\widehat D}_t{{\cal P}})\,=\,0\ .\]
\end{theorem}
This result can be called the {\it ``Noether $\Lambda^{({{\cal L}})}$-conservation 
rule''}. Indeed, if $\Lambda^{({{\cal L}})}\!=\!0$, the standard Noether 
theorem is recovered.

\smallskip
In the special case $\Lambda^{({{\cal L}})}\varphi=\lambda \varphi$,  
the above result becomes
\[{\widehat D}_t{\bf P}=0\quad\quad {\rm where}\quad\quad  
{\widehat D}_t=D_t+\lambda\ .\]

Theorem 3 can be extended \cite{Cic:CGNoe} to divergence symmetries and to 
generalized symmetries as well. Also, higher-order Lagrangians can  be included:
the $\Lambda^{({{\cal L}})}$-conservation rule has the 
same form, but ${{\cal P}}_{\alpha\beta}$ is different: for instance, for 
second-order Lagrangians one has
\[{{\cal P}}_{\alpha\beta}\,=\, 
\varphi_\alpha{\frac{\partial {{\cal L}}}{\partial \dot q_\beta}} + 
(({\widehat D}_t)_{\alpha\gamma}  \varphi_\gamma) 
{\frac{\partial {{\cal L}}}{\partial 
\ddot q_\beta}} - \varphi_\alpha D_t{\frac{\partial {{\cal L}}} 
{\partial\ddot q_\beta}} \ .\]

\subsection{From Lagrangians to Hamiltonians}

Assume to have a Lagrangian which is $\Lambda^{({{\cal L}})}$-invariant 
under a vector field 
\[X^{({{\cal L}})}\,=\,\varphi_\alpha{\frac\partial {\partial q_\alpha}}\]
and introduce the corresponding Hamiltonian $H$ with its Hamiltonian 
equations of motion. The natural question is whether the 
$\Lambda^{({{\cal L}})}$-symmetry $X^{({{\cal L}})}$ of the Lagrangian is 
transferred to some
$\Lambda^{(H)}$-symmetry $X^{(H)}$ of the Hamiltonian equations of motion.
Two problems then arise:  $i)$ to extend the vector field $X^{({{\cal L}})}$ to 
a suitable vector field $X^{(H)}$, and $ii)$ to extend the $(n\times n)$ matrix 
$\Lambda^{({{\cal L}})}$ to a suitable $(2n\times 2n)$ matrix 
$\Lambda^{(H)}$.

First, the vector field $X^{(H)}$ is expected to have the form
\begin{equation}\label{Cic:Lapq} X\equiv X^{(H)}\,=\,
\varphi_\alpha{\frac\partial {\partial q_\alpha}}+
\psi_\alpha{\frac\partial {\partial p_\alpha}} 
\end{equation}
where the coefficient functions $\psi$ must be determined.
This can be done observing that the variables $p$ are related to $\dot q$ (and then
the first $\Lambda^{({{\cal L}})}$-prolongation of $X^{({{\cal L}})}$ is needed, 
where the ``effect'' of $\Lambda^{({{\cal L}})}$ is present). 
One finds, after some explicit calculations,
\begin{equation}\label{Cic:XH}
\psi_\alpha\,=\,{\frac\partial {\partial \dot q_\alpha}}\Big(D_t  
{\bf P}+\Lambda^{({{\cal L}})}_{\beta\gamma}\varphi_\gamma{\frac{\partial 
{{\cal L}}}{\partial\dot q_\beta}}\Big)-{\frac{\partial\Lambda^{({{\cal L}})}_
{\beta\gamma}}{\partial\dot q_\alpha}}\varphi_\gamma{\frac{\partial{{\cal L}}}
{\partial\dot q_\beta}}-
p_\beta{\frac{\partial\varphi_\beta}{\partial q_\alpha}}\ .
\end{equation}
But the term in parenthesis vanishes if the Lagrangian is 
$\Lambda^{({{\cal L}})}$-invariant, thanks to Theorem 3; in addition,
if $\Lambda^{({{\cal L}})}$ does not depend on $\dot q$ (as happens in most cases, 
otherwise a
separate treatment is needed, see subsect. 3.4),  then we are left with
\begin{equation}\label{Cic:psiH}
\psi_\alpha=-p_\beta{\frac{\partial\varphi_\beta}{\partial q_\alpha}}\ . 
\end{equation}
This implies that $X$ admits a  generating function, which is just
\[G\,=\,\varphi_\alpha p_\alpha\equiv{\bf P} \]
using the  notations introduced in Theorem 3.

Second, let me now introduce the following $(2n\times 2n)$ matrix 
\[\Lambda\equiv\Lambda^{(H)}\,=\,\pmatrix{\Lambda^{({{\cal L}})} & 0\cr  -
{\frac{\partial\Lambda^{({{\cal L}})}}{\partial 
q_\alpha}}p_\gamma & \Lambda^{(2)}} \]
where  $\Lambda^{(2)}$ must satisfy 
($\Lambda$ is not uniquely defined, as well known)
\[\Lambda^{(2)}_{\alpha\beta}\,{\frac{\partial\varphi_\gamma}{\partial
q_\beta} }\,=\,\Lambda^{({{\cal L}})}_{\gamma\beta}
{\frac{\partial\varphi_\beta} {\partial q_\alpha}}\ .\]
It is well known that Euler-Lagrange equations coming from a
$\Lambda^{({{\cal L}})}$-invariant Lagrangian do {\it not  exhibit in general 
$\Lambda$-symmetry}. In contrast with this, it is not difficult to verify 
explicitly that the Hamiltonian equations of motion turn out to be 
$\Lambda^{({H)}}$-symmetric under the vector field $X^{({H)}}$ obtained  
according to the above prescriptions. 

\smallskip
In conclusion, I have shown the following 

\begin{theorem}
If ${{\cal L}}$ is a  $\Lambda$-invariant Lagrangian under a vector 
field $X^{({{\cal L}})}$ 
with a matrix $\Lambda^{({{\cal L}})}$ (not depending on $\dot q$), one can 
extend $X^{({{\cal L}})}$ to a vector field\ 
$X\equiv X^{(H)}$ and the $(n\times n)$ matrix $\Lambda^{({{\cal L}})}$ 
to a $(2n\times 2n)$ matrix $\Lambda\equiv \Lambda^{(H)}$ in 
such a way that  the resulting Hamiltonian equations of motion 
are $\Lambda$-symmetric under $X$; in addition,  $G=\varphi_\alpha p_\alpha$ 
is a $\Lambda$-constant of motion.
\end{theorem}

\smallskip\smallskip
\begin{example}
The Lagrangian (with $n=2$)
\[{{\cal L}}={\frac 1  2}\Big({\frac{\dot q_1} {q_1}}-q_1\Big)^2+{\frac 1 
2}(\dot q_1-q_1\dot q_2)^2\exp(-2q_2)+q_1\exp(-q_2) \]
is $\Lambda^{({{\cal L}})}$-invariant under 
\[X^{({{\cal L}})}\,=\,q_1{\frac\partial {\partial q_1}}+{\frac\partial 
{\partial q_2}}\]
with
\[\Lambda^{({{\cal L}})}\,=\,{\rm diag}\ (q_1,q_1)\ .\]
It is easy to write the Hamiltonian equations of motion and to check that 
they are indeed $\Lambda$-symmetric under
\[ X\,=\,q_1{\frac\partial {\partial q_1}}+
{\frac\partial {\partial q_2}}-p_1{\frac\partial {\partial p_1}}\]
with    
\[ \Lambda\,=\,\Lambda^{(H)}\,=\,\pmatrix {q_1&0&0&0\cr 0&q_1&0&0\cr
-p_1&-p_2&q_1&0\cr 0&0&0&0}\ . \]
$X$-invariant
coordinates are $w_1=q_1\exp(-q_2),\,w_2=q_1p_1,\,w_3=p_2$, 
and, as expected, the generating function $G=w_2+w_3$ satisfies the 
$\Lambda$-conservation rule 
\[\nabla_uD_tG\,=\,J\Lambda{\bf \Phi} \quad\ {\rm or}\quad D_tG\,=\,-q_1G\ .\]
\end{example}

\smallskip
A special, but rather common, case is described by the following:

\begin{corollary}
If \[\Lambda^{({{\cal L}})}{\bf \varphi}=c\, \varphi\]  where $c$ is a constant, 
then also $\Lambda{\bf \Phi}=c\,{\bf \Phi}$ and the ``most
complete'' reduction  of the Hamiltonian equations  
of motion  is obtained:
\[\dot G=\gamma(G,t)\quad\quad \dot w_j=W_j(w,G,t)\quad\quad \dot z=Z(w,G,z,t)\]
\end{corollary}

\subsection{Reduction of the Euler-Lagrange equations versus\\ 
the Hamiltonian equations}

In this section I want to compare the reduction procedure which is provided 
by the presence of a $\Lambda$-symmetry of a  Lagrangian (i.e. the reduction  
of Euler-Lagrange equations) with the analogous reduction of the Hamiltonian 
equations of motion.
 
Let me start recalling that any vector field\
$X=\varphi_\alpha\partial/\partial q_\alpha$ admits 
$n$ (0-order) invariants (as already said, see subsect. 2.1)
\[w_j=w_j(q,t)\quad\ (j=1,\ldots,n-1)  \quad {\rm and \ the\ time}\ t\] 
and $n$   
first-order differential invariants $\eta_\alpha=\eta_\alpha(q,t,\dot q)$ under 
the first prolongation $X^{(1)}$
\[X^{(1)}\eta_\alpha\,=\,0 \quad\quad\quad\ (\alpha=1,\ldots,n)\ .\]
{\it Both} if $X^{(1)}$ is standard  and if it is a $\Lambda$ prolongation 
({\it under the condition} $\Lambda\,\varphi=\lambda\,\varphi$),
it is well known that
$\dot w_j$ are $n-1$ first-order differential invariants
(notice that this is an 
``algebraic'' property, not related to  dynamics).
If one now chooses another independent first-order differential invariant 
$\zeta=\zeta(q,t,\dot q)$, 
then one has that any  first-order $\Lambda^{({{\cal L}})}$-invariant
Lagrangian  is a function of the above $2n$ invariants 
\[t,w_j,\dot w_j\quad{\rm and}\quad  \zeta \ .\]
Writing the Lagrangian in terms of these variables, the Euler-Lagrange 
equation for $\zeta$ is then simply
\[  {\frac{\partial  {{\cal L}}}{\partial\zeta}}\,=\,0\ .\]
This  first-order equation   provides in general a ``partial'' 
reduction, i.e.,  it produces  only   {\it particular  solutions}, 
even  considering the  Euler-Lagrange equations for the other variables 
\cite{Cic:MRO,Cic:CHam}
(notice that this is true both for exactly invariant and for 
$\Lambda^{({{\cal L}})}$-invariant Lagrangians).

\smallskip
I want to emphasize that, introducing   $\Lambda$-symmetric Hamiltonian 
equations of motion along the lines stated in Theorem 4,  then a  ``better'' reduction is 
obtained, and no solution is lost. The following example clarifies this point.

\smallskip
\begin{example}
The Lagrangian ($n=2$)
\[{{\cal L}}\,=\,{\frac 1  2}\Big({\frac{\dot q_1}{q_1}}-\log q_1\Big)^2+{\frac 1  
2}\Big({\frac{\dot q_1}{q_1}}+{\frac{\dot q_2}{q_2}}\Big)^2\quad\quad (q_1>0)\]
is $\Lambda^{({{\cal L}})}$-invariant under
\[X^{({{\cal L}})}\,=\,q_1{\frac\partial {\partial q_1}}-q_2{\frac\partial 
{\partial q_2}}\]
with $\Lambda^{({{\cal L}})}={\rm diag}\ (1,1)$. With
\[ w\,=\,q_1q_2,\ \dot w\,=\,\dot q_1q_2+q_1\dot q_2,\ \zeta\,=\,{\frac{\dot q_1}{ q_1}}-\log 
q_1 \]
the Lagrangian becomes
\[  {{\cal L}}\,=\,{\frac 1 2}\zeta^2+{\frac1 2}{\frac{\dot w^2}{w^2}}\]
and the Euler-Lagrange equation for $\zeta$ is
\[{\partial  {{\cal L}}/{\partial \zeta}}=\zeta=0\quad\quad\ {\rm or}\quad\
\dot q_1\,=\,q_1\log q_1\] 
with the particular solution
\[q_1(t)\,=\,\exp(c\, e^t)\ .\]
The corresponding
Hamiltonian equations of motion are $\Lambda$-symmetric under 
\[ X\,=\, q_1{\frac\partial {\partial q_1}}-
q_2{\frac\partial {\partial q_2}}-p_1{\frac\partial {\partial 
p_1}}+p_2{\frac\partial {\partial p_2}}\]
with $\Lambda={\rm diag}\ (1,1,1,1)$.
Invariants under this $X$ are
\[w_1\,=\,q_1q_2,\ w_2\,=\, q_1p_1,\ w_3\,=\,q_2p_2\]
and $X$ is generated by $G=w_2-w_3$. A ``complete'' reduction is obtained: 
with  $z=\log q_1$, we get
\[\dot w_1\,=\,w_1w_3 \quad\quad \dot w_2\,=\,w_3-w_2 \] 
\[ \dot G\,=\,-G \quad\quad\ \dot z\,=\,z+w_2-w_3\]
The above ``partial'' (Lagrangian) solution $\zeta=0$ 
corresponds~to 
\[\dot z\,=\,z,\quad w_2=w_3=c= {\rm const},\quad  \dot w_1=cw_1\ .\] 
From the Hamiltonian equations, instead, e.g.:
\[q_1(t)\,=\,\exp(c\, e^t)+c_1\exp(-t)\quad\quad {\rm etc.}\]
The reader can easily complete the calculations.
\end{example}

\subsection{When $\Lambda^{({{\cal L}})}$ depends on $\dot q$}

If $\Lambda$ depends also on $\dot q$ (see eq.s (\ref{Cic:Lapq},\ref{Cic:XH})),
the calculations performed in subsect. 3.2 cannot be repeated, the 
coefficient functions $\psi_\alpha$ cannot be expressed in the simple 
form (\ref{Cic:psiH}) and the vector field  
$X$ does not admit a generating function $G$. In this case one can resort to 
the other quantity $S$, introduced in (\ref{Cic:defS}), which provides a 
$\Lambda$-constant of motion. An example can completely illustrate this situation.

\begin{example}
($n=1$)
\[{{\cal L}}\,=\,{\frac 1  2}\Big({\frac{\dot q}{q}}+1\Big)^2\exp(-2q)\]
is $\Lambda^{({{\cal L}})}$-invariant under
\[X^{({{\cal L}})}\,=\,q{\frac\partial {\partial q}}\quad\quad\ 
{\rm with}\quad\quad \Lambda^{({{\cal L}})}\,=\,q+\dot q\ .\]
One finds $\psi=-qp-p$ 
and the resulting vector field
\[ X\,=\,q{\frac\partial {\partial q}}-(qp+p){\frac\partial {\partial p}}\]
does {\it not} admit a generating function. Nevertheless, the Hamiltonian 
equations of motion are $\Lambda$-symmetric under $X$ with
\[\Lambda\,=\,\pmatrix{
q+\dot q & 0 \cr -p & q+\dot q}\ .\]
Here
\[S\,=\,-q\]
satisfies $D_tS=-\nabla(\Lambda\Phi)$  and is  a $\Lambda$-constant of motion. 
\end{example}

\section{A digression: general $\Lambda$-invariant Lagrangians}

The $\Lambda$-invariance of a Lagrangian  
${{\cal L}}={{\cal L}}(q,\dot q,t)$ considered in subsect.~3.1 is a special case of 
a much more general situation.
Instead of $n$ time-dependent quantities $q_\alpha(t)$, let me consider 
now $n$ ``fields'' 
\[u_\alpha(x_i)\quad (\alpha=1,\ldots,n\,;\,i=1,\ldots,s)\]
depending on $s>1$ real variables $x_i$.
Now, the Euler-Lagrange equations become a system of PDE's, and 
the notion of $\mu$-symmetry \cite{Cic:GM,Cic:CGMmu} extends and replaces that of 
$\lambda$-symmetry (or $\Lambda$-symmetry if $n>1$).

\smallskip
In this case, there are $s>1$ matrices $\Lambda_i$ ($n\times n$), 
which must satisfy the compatibility condition
\begin{equation}\label{Cic:LaLa}D_i\Lambda_j-D_j\Lambda_i+[\Lambda_i,\,
\Lambda_j]\,=\,0\quad \quad (D_i\equiv D_{x_i}) \end{equation}
which can be rewritten putting ${\widehat D}_i\,=\,D_i\,\delta+\Lambda_i$ 
(or, in explicit form: 
$({\widehat D}_i)_{\alpha\beta}\,=\,D_i\delta_{\alpha\beta}+
(\Lambda_i)_{\alpha\beta}$, with a notation 
extending the one introduced in Theorem 3),
\[  [ {\widehat D}_i , {\widehat D}_j ] \ = \ 0 \ .\]
Then one has \cite{Cic:CGNoe,Cic:GM,Cic:CGMmu}:

\begin{theorem} 

Given $s>1$ matrices $\Lambda_i$ satisfying (\ref{Cic:LaLa}), there exists 
(locally) a $(n\times n)$  nonsingular matrix $\Gamma$ such that
\[\Lambda_i\,=\,\Gamma^{-1}(D_i\Gamma)\ .\]
If a Lagrangian ${{\cal L}}$ is $\Lambda$-invariant under a vector field
\[ X\,=\,\varphi_\alpha{\frac {\partial} {\partial u_\alpha}}\]
then there is a matrix-valued vector 
\[{{\cal P}}_i\equiv({{\cal P}}_i)_{\alpha\beta}\] 
which is $\Lambda$-conserved; this $\Lambda$-conservation law holds in the form
\[
{\tt {Tr}}\, \big[ \Gamma^{-1}D_i\big( \Gamma\,{{\cal P}}_i \big)\big] 
\,=\,  0  \]
or in the equivalent forms
\[ 
D_i{{\bf P}}_i\,=\,-(\Lambda_i)_{\alpha\beta}({{\cal P}}_i)_{\beta\alpha} 
\,=\,- {\tt {Tr}}(\Lambda_i{{\cal P}}_i)\  ,\ \  where \quad 
{{\bf P}}_i=({{\cal P}}_i)_{\alpha\alpha}\,=\,{\tt Tr}\,{{\cal P}}_i \ ,\]
\[{\tt {Tr}}({\widehat D}_i\,{{\cal P}}_i)\,=\,0\ .\]
\end{theorem}

\smallskip\noindent
For first-order Lagrangians the $\Lambda$-conserved ``current density 
vector'' ${\cal P}_i$ is given by
\[({{\cal P}}_i)_{\alpha\beta}\,=\,\varphi_\alpha{\frac{\partial {\cal L}}
{\partial u_{\beta,i}}} \quad\quad \quad  {\rm where}\quad
\quad u_{\beta,i}\,=\,{\frac{\partial u_\beta} {\partial x_i}}\]
and for second-order Lagrangians  by
\[ ({{\cal P}}_i)_{\alpha\beta}\,=\,
\varphi_\beta{\frac {\partial{{\cal L}}}{\partial u_{\alpha,i}}}+
(({\widehat D_j)_{\beta \gamma}\varphi_\gamma){\frac{\partial {{\cal L}}}
{\partial u_{\alpha,ij}}}-\varphi_\beta D_j {\frac{\partial {{\cal L}}}
{\partial u_{\alpha,ij}}}}\ .\]

\medskip

\begin{example} Let $n=s=2$. Writing for ease of notation, $x,y$ instead 
of $x_1,\, x_2$, 
and $u=u(x,y), \,v=v(x,y)$ instead of $u_1,\,u_2$, consider the vector 
field
\begin{equation}\label{Cic:exe4}X\,=\,u{\frac \partial {\partial u}}+
{\frac \partial {\partial v}}\end{equation}
and the two matrices
\[\Lambda_1\,=\,\pmatrix{0 & 0\cr u_x  & 0} \quad\quad \Lambda_2\,=
\pmatrix { 0 & 0\cr u_y  & 0 } \] 
and then
\[\Gamma\,=\,\pmatrix{1 & 0 \cr u & 1}\ .\]
It is easy to check that the Lagrangian
\[{\cal L}\,=\, {\frac 1 2}\Big( u_x^2+u_y^2\Big)-{\frac 1
u} \big( u_xv_x+u_yv_y \big) + u^2 \exp(-2v) \]
is $\Lambda$-invariant (or better, in this context, $\mu$-invariant)
but not invariant   under the above vector field  $X$. 
The $\mu$-conservation law ${\tt Tr} (\widehat D_i {{\cal P}}_i)=0$ takes 
here the form
\[ D_i  {\bf P}^i \equiv D_x \big( uu_x-v_x-{\frac {u_x} u} \big)+  
D_y\big(uu_y-v_y-{\frac {u_y} u}\big)\,=\,  u_x^2+u_y^2\ .\]
In agreement with Theorem 5, the r.h.s. of  this expression is precisely
equal to
\[-{\tt Tr}(\Lambda_i{\cal P}_i)=
-(\Lambda_i\varphi)_\alpha{\frac {\partial {\cal L}} {\partial u_{\alpha,i}}}\ .\] 
Notice that in this case the quantity $u_x^2+u_y^2$ is just the 
``symmetry-breaking term", 
i.e. the term which prevents the above Lagrangian from being exactly symmetric
under  the vector field (\ref{Cic:exe4}). 
\end{example}

\smallskip

It should be remarked that $\mu$-symmetries are actually strictly related to 
{\it standard} symmetries, or -- more precisely -- are {\it locally 
gauge-equivalent} to them (see for details \cite{Cic:CGNoe,Cic:Gatw,Cic:Ggau}).

Given indeed the vector field $X=\varphi_\alpha\partial/\partial u_\alpha$ 
and the $s$ matrices $\Lambda_i$, let me denote by
\[X_\Lambda^{(\infty)}\,=\,\sum_J\Psi^{(J)}_ \alpha {\frac \partial 
{\partial u_{\alpha,J}}}\]
the infinite $\Lambda$-prolongation of $X$, where the sum is over all 
multi-indices $J$ as usual, and $\Psi^{(0)}_\alpha=\varphi_\alpha$. Introducing 
now the other vector field $\widetilde X$
\[
\widetilde X\equiv\widetilde\phi_\alpha{\frac\partial{\partial u_\alpha}}\  \quad {\rm with}\quad
\ \widetilde\varphi_\alpha\equiv(\Gamma\,\varphi)_\alpha \] 
where $\Gamma$ is assigned in Theorem 5, and denoting by
\[\widetilde X^{(\infty)}\,=\,
\sum_J\widetilde\varphi^{(J)}_\alpha{\frac \partial 
{\partial u_{\alpha,J}}}\]
the {\it standard} prolongation of $\widetilde X$, 
one has \cite{Cic:CGMmu,Cic:CGNoe} that the coefficient functions $\Psi^{(J)}_\alpha$ 
of the $\Lambda$ 
prolongation of $X$ are connected to the coefficient functions 
$\widetilde\varphi^{(J)}_\alpha$ 
of the standard prolongation of $\widetilde X$ by the relation
\[\Psi^{(J)}_\alpha\,=\,\Gamma^{-1}\widetilde\varphi^{(J)}_\alpha \ .\]
In the particularly simple case $n=1$ (i.e., a single ``field" 
$u(x_i)$), then the $s>1$ matrices $\Lambda_i$, and  the matrix $\Gamma$ 
as well, become (scalar) functions $\lambda_i$ and $\gamma$; in this case,
if a Lagrangian is
$\mu$-invariant under the vector field $X$, then it is also invariant 
under the {\it standard} symmetry $\widetilde X=\gamma X$. In addition, the $\mu$ 
conservation law can be  also expressed as a {\rm standard} conservation rule
\[ D_i{\bf\widetilde P}^i\,=\,0 \] 
where ${\bf\widetilde P}^i=\gamma\,\varphi_\alpha\partial{\cal L}/\partial 
u_{\alpha,i}$ is the ``current density
vector'' determined by the vector field $\widetilde X=\gamma X$.

\begin{example}
Let now $n=1,\,s=2$, and let me introduce for convenience as independent variables
the polar coordinates $r,\theta$.  I am considering a single ``field"
$u=u(r,\theta)$ and the rotation vector field  $X=\partial/\partial\theta$.
The Lagrangian
\[{\cal L}\,=\, {\frac 1 2} r^2\exp(-\epsilon\theta)u_r^2+{\frac 1 2} 
\exp(\epsilon\theta)u_\theta^2\]
is clearly not invariant under rotation symmetry (if
$\epsilon\not=0$), but is $\mu$-invariant with 
$\lambda_1=0,\,\lambda_2=\epsilon$.
The above Lagrangian is the Lagrangian of a perturbed Laplace equation, indeed the 
Euler-Lagrange equation is the PDE
\[r^2u_{rr}+2ru_r+\exp(2\epsilon\theta)(u_{\theta\theta}+\epsilon\, 
u_\theta)\,=\,0 \ .\]
It is easy to check that the current density vector
\[{\bf P}\equiv\big(-r^2\exp(-\epsilon\theta)u_ru_\theta\, , \, {\frac 1 2}
r^2\exp(-\epsilon\theta)u_r^2-{\frac 1 2} 
\exp(\epsilon\theta)u_\theta^2\big)\]
satisfies the $\mu$-conservation law
\[D_i{\bf P}_i\,=\,-\epsilon {\bf P}_2 \ .\]
According to the above remark on  the (local) equivalence of the 
$\mu$-symmetry $X$ to the standard symmetry 
$\widetilde X\,=\,\gamma\,X\,=\,\exp(\epsilon\theta)\,{\partial/\partial\theta}$, 
also the 
(standard) conservation law $D_i\widetilde{{\bf P}}^i=0$ holds, with 
\[\widetilde{{\bf P}}\equiv \Big(-r^2 u_ru_\theta\, , \, {\frac 1 2} r^2 u_r^2-
{\frac 1 2}\exp(2\epsilon\theta)u_\theta^2\Big)\ .\]
\end{example}

\section{Conclusions}
I have shown that the notion of $\lambda$-symmetry, and the related procedures for studying differential equations, can be conveniently 
extended to the case of dynamical systems. 

The use and the interpretation of this 
notion becomes particularly relevant when the DS is a Hamiltonian system, 
and even more if the symmetry is inherited by an invariant Lagrangian:
in this context indeed it is possible to introduce in a natural way and to draw a comparison 
between the notions of  $\Lambda$-constant of 
motion and of   Noether $\Lambda$-conservation rule.
Similarly, the symmetry properties of Euler-Lagrange equations and 
of the Hamiltonian ones can be compared, and some reduction techniques for the equations can 
be conveniently introduced. 

Finally, I have shown that the $\Lambda$-invariance
of the Lagrangians in the context of the DS is a special case of a more 
general and richer situation, where several independent variables are present and a 
$\Lambda$-conservation rule of very general form is true.

Another interesting problem is the nontrivial relationship between $\lambda$ (or $\Lambda$, or $\mu$) symmetries with the standard ones. An aspect of this problem has been mentioned in the above section of this paper. In different situations, this may involve the introduction of nonlocal symmetries and other concepts in differential geometry, as briefly indicated in the Introduction, which clearly go beyond the scope of the present contribution. 


\end{document}